\documentstyle[prb,preprint,aps]{revtex}
\input epsf

\tighten

\begin{document}
\draft

\title{Quantum and Many-Body Effects on the\\
       Capacitance of a Quantum Dot}


\author{
          Lotfi Belkhir \\[10pt]
          Department of Physics,\\
          State University of New York, \\
          Stony Brook, New York 11794-3800 \\[15pt]
         }

\maketitle

\begin{abstract}

We calculate exactly, using finite size techniques, the
quantum mechanical and many-body effects to the
self-capacitance of a spherical quantum dot
in the regime of extreme confinement, where the radius of the sphere
is much smaller than the effective Bohr radius.
We find that the self capacitance oscillates as a
function of the number of electrons close to its classical value.
We also find that the electrostatic energy as a function
of the number of electrons extrapolates to zero when $N=1$, suggesting
that the energy scales like $e^{2}N(N-1)$ instead of $(N\,e)^2$.
We also provide evidence that the main deviations from the
semiclassical description are due to the exchange interaction
between electrons.
This establishes, at least for this configuration, that the
semiclassical description of Coulomb charging effects in terms of
capacitances holds to a good approximation even at very small scales.

\end{abstract}


\mediumtext

With the rapid advances in the fabrication of increasingly smaller
quantum dots, approaching the atomic scale, the question of
quantum and many-body effects on the essential characteristics
of these objects has become a central issue.
The current stage of theoretical understanding of quantum dots
relies essentially on a semi-classical picture\cite{AL,HBS},
which makes the assumption
that the Coulomb charging effects can be described in terms of
classical capacitances, where the resonance energies
can be separated in a single-particle confinement energy, and a
constant Coulomb charging energy terms.
The subject has been further studied in many recent
theoretical\cite{Bryant,GS,KLS,MWL,GIV,stopa} and experimental
investigations\cite{MKW,Kal,Mal,Aal,GSC}.
The classical picture of Coulomb blockade, however, has recently been
questioned by Johnson and Payne\cite{JP2} who,
using a harmonic model
interaction that is exactly solvable\cite{JP1},
showed that, in presence of magnetic field, the resonance
energies could not be written as the sum of single-particle and
charging energies terms. They argued that
the model interaction shows a behavior similar to a Coulomb
interaction with a cutoff\cite{Banyai} for a certain range of
electron-electron
separation. On the other hand, however, the semiclassical description
seems so far to provide a qualitatively correct picture, given that
some experiments\cite{McEuen,GSC}, performed in the regime where
confinement and charging energies are of equal importance, can be
well explained by this model.

\noindent
We investigate this issue further, using exact finite
size calculations techniques. We consider an isolated spherical
quantum dot in zero magnetic field, and solve numerically the
full Coulomb interaction problem, for up to 30 electrons, in the
regime where the quantum effects are expected to be maximal,
i.e when the radius of the dot is much smaller
than the effective Bohr radius (i.e $R\ll a_0$).
It is assumed that the added electrons move on the surface of the
sphere, which could be a reasonable model for a metallic sphere.
It has recently been shown\cite{mhi}, in the framework of
density-functional theory, that in most experimental situations
the main contribution to the
capacitance of a quantum dot, in the presence of leads and backgates,
comes from the self-capacitance. The contributions of the leads and
backgates was found to be $30\%$ at most. The results for an isolated
dot are therefore not irrelevant to actual experiments.

\noindent
We do the calculation for both spin unpolarized and spin polarized
cases. We find that:
(i) the interaction
energy spectrum scales like $N(N-1)e^2/2C$, where $N$ is the number
of electrons at the surface of the dot, and $R$ its radius.
This corroborates the semi-classical expression for the charging
energy suggested recently by some authors\cite{Lent,AKL}, instead
of the more widely used expression $(Ne)^2/2C$.
(ii) the resonance energies  {\sl do} separate into confinement
and charging energies to a good approximation;
(iii) the self-capacitance of the isolated
dot, defined from the charging energy, oscillates as a function
of $N$, around its classical value, i.e $R$, and gets closer to
$R$ as $N$ increases.
(iv) The main deviation of the self-capacitance from the classical value
are due to the exchange interaction between electrons, which, in
agreement with Hund's first rule, tend to
lower the ground state energy of the system, and thus increase the
value of the capacitance. This increase is however never greater than
$25\%$ for $N>2$.

\noindent
Our Hamiltonian for an isolated spherical dot, is given by

\begin{equation}
   H = \frac{1}{2m^*R^2}\sum_{i}^{N}|{\bf L}_{i}|^{2} +
       \sum_{i<j}\frac{e^2}{\epsilon|{\bf r}_i -  {\bf r}_j|},
\end{equation}
where ${\bf L}_{i}=-i\hbar{\bf R}_i\times{\bf \nabla}_i$ and
      ${\bf r}_{i}$
are the angular momentum and the
position of the $i$-th particle, and $\epsilon$ the dielectric
constant. The eigenvalues of $|{\bf L}|^{2}$ are equal to
$l(l+1)$ with $l$ an integer. The kinetic energy of an electron
in the shell $l$ is $\varepsilon_l=\hbar^2/(2m^*R^2)~l(l+1)$.
The maximum number of electrons
in a shell of angular momentum $l$ is $2l+1$ and $2(2l+1)$
for a spin polarized and spin unpolarized cases respectively.

\noindent
The dot contains $N$ electrons, with an effective mass
$m^*$, and charge $-|e|$.
Notice the absence of a confinement
term, due to the use of a spherical geometry, where the electrons
are constrained to move on the surface of the sphere of constant
radius $R$. Previous  quantum mechanical calculations of the capacitance
of quantum dots\cite{stopa,mhi} involved disc geometries with
parabolic confinement potential. It was found in those calculations
that the effective size of the dot increases with the number of
electrons. in our case however, the size of the dot is absolutely
rigid, which will facilitate considerably the comparison of our
calculations with the classical results.

\noindent
The Coulomb
interaction between two electrons moving on the surface of a sphere
with radius $R$ can be rewritten as

\begin{eqnarray}
V({\hat \Omega}_1, {\hat \Omega}_2) & = & \frac{e^2}{R}
                \frac{1}{|{\hat \Omega}_1 - {\hat \Omega}_2|}\nonumber\\
  & =  & \frac{e^2}{R}\sum_{l,m}(\frac{4\pi}{2l+1})~
      Y_{lm}^*(\Omega_1) Y_{lm}(\Omega_2),
\end{eqnarray}

\noindent
In second quantization form, the interaction operator is given by
\begin{equation}
\hat{V} = \frac{1}{2}\sum_{\sigma_1,\sigma_2}\int~d\Omega_1~d\Omega_2
 \Psi_{\sigma_1}^{\dagger}({\bf \Omega_1})
\Psi_{\sigma_2}^{\dagger}({\bf\Omega_2})
 V({\bf \Omega}_1, {\bf \Omega}_2)
 \Psi_{\sigma_2}({\bf \Omega_2})\Psi_{\sigma_1}({\bf \Omega_1}),
\end{equation}
where the $Y_{lm}^*(\hat \Omega)$ are the usual spherical harmonics, and
$\Psi_{\sigma}^{\dagger}(\hat \Omega) = \sum_{l,m}Y_{lm}(\hat \Omega)~
a_{lm\sigma}^{\dagger}$.

\noindent
To carry out our numerical calculations, we choose the convenient
single-particle basis states defined by

\begin{equation}
<\Omega|lm\sigma> = Y_{lm}(\Omega)~\chi_{\sigma}.
\end{equation}

\noindent
In this basis the two body interaction operator is given by
\begin{equation}
\hat{V} = \sum_{all indices} \frac{1}{2} v_{_{l_1m_1l_2m_2l_3m_3l_4m_4}}
c^{\dagger}_{_{l_2m_2\sigma}}
c^{\dagger}_{_{l_3m_3\sigma^{\prime}}}
c_{_{l_4m_4\sigma^{\prime}}}
c_{_{l_1m_1\sigma}},
\end{equation}
where the matrix elements are given by
\begin{eqnarray}
v_{_{l_1m_1l_2m_2l_3m_3l_4m_4}} &=& \sum_{L,M}~(-1)^{l_1+l_4-l_2-l_3}
\left[\frac{(2l_1+1)(2l_4+1)}{(2l_2+1)(2l_3+1)}\right]
 <l1,m1;L,M|l_3,m_3>\\
& &<l_1,0;L,0|l_3,0> <l_4,m_4;L,M|l_2,m_2>~<l_4,0;L,0|l_2,0>
\end{eqnarray}
where the terms $<l,m;l_1,m_1|l_2,m_2>$ are the usual Clebsh-Gordon
coefficients, which are non zero only when $m+m_1-m_2=0$.

\noindent
We also assume the limit of $R\rightarrow\,0$, when the energy
separation between successive angular momentum shells is large
compared to the Coulomb interaction energy, so that mixing between
shells can be neglected. This is equivalent to assuming that
$R\ll\,a_0$, where $a_0=\frac{\epsilon\hbar^2}{m^*e^2}$ is the
effective Bohr radius, determined solely by the material's properties.
In GaAs quantum dots, $a_0=10nm$, and the confinement energy
$\hbar^2/(2m^*R^2)$ is typically about 15/,meV, which for a spherical
dot yields a radius $R=8nm$. The strong confinement regime could be
attained by either reducing the size of the dots, or using materials
with a higher dielectric constant $\epsilon$, which would increase
the effective Bohr radius $a_0$.
The assumption of $R\rightarrow\,0$ drastically
reduces the Hilbert size of the quantum system, since intershell
transitions can be completely ignored, which allows us to do the
calculations for up to 30 electrons. The calculation done in this
limit is similar to the finite size calculations done in the context
of the fractional quantum Hall effect\cite{haldane}, where the limit
of infinite magnetic field is assumed in order to ignore transitions
to higher Landau levels.

\noindent
We calculate the energy spectrum by exact diagonalization of the
Hamiltonian.
The calculation is done by filling the angular momentum
shells  by adding electrons one by one, and calculating the ground
state of the whole many-body system.
In less than half-filled
shell, all electrons tend to have
the same spin in order to gain the exchange energy in accordance
with Hunds' rule. As half-filling is reached, there is
a sudden increase in the interaction energy due to the fact that
the additional electrons must have opposite spins, in order to
satisfy the Pauli principle, and thus lose the exchange energy.
Our results for the ground state interaction energy $E_c(N)$
as a function of $N$, spin unpolarized case, are
shown in figure 1. It shows an almost linear behavior, and
extrapolates to $0$ when $N=1$, which is
consistent with the semiclassical expression $N(N-1)e^2/R$.

\noindent
We first define the chemical potential as
\begin{equation}
    \mu(N) = E_c(N) - E_c(N-1)
\end{equation}
\noindent
We define the self-capacitance of our dot as
$C = \Delta\,Q/\Delta\,V$  which can be readily obtained
from the chemical potential. For a single electron $\Delta\,Q = e$
and $\Delta\,V = [\mu(N+1) - \mu(N)]/e$.
Thus
\begin{equation}
     C_{dot}(N) = e^2/[\mu(N+1) - \mu(N)]
\end{equation}
In the case where tunneling occurs through the channel where
the dot is at its ground state before and {\sl after} the event,
the resonance energies are identical to the chemical potential.

\noindent
Notice that the definition of the chemical potential is usually
taken as the difference between the total energies, rather than
the interaction energies. In our case, both definitions are
identical for electrons in the same shell, since the kinetic
energy within one shell is constant. The only difference occurs
when $N$ electrons correspond to a state of completely filled
shells, and the $(N+1)$th electron must occupy the next empty shell.
The jump in the chemical potential will then be incremented
by the confinement energy, in addition to the charging energy.
Since we are choosing the confinement energy to be extremely
large, we always subtract it from the total energy to keep only
the interaction part.

\noindent
Figure~(2a) and (2b) show the numerical results for the chemical
potential, and the self-capacitance
as a function of $N$ in units of $R$ for a spin unpolarized electron
system. We see clearly that the
capacitance is quite close to its classical value, modulo some
quantum fluctuations. The peaks observed are due to the sudden increase
in energy that occurs at half filling of each angular momentum shell.
Notice also that, except for half filled shells, the quantum and
many-body effects tend to increase the capacitance only slightly . The
average over all the 30 electrons, including the values at the
peaks is $1.117R$. Figure~(3a) and (3b) show the results for the
chemical potential and the self-capacitance of the spin polarized
case.
Observe now how the capacitance has became consistently greater
than its classical value. However the difference is not
greater than $25\%$ for all $N\leq~30$. The average capacitance for
the spin polarized case over $N_{max}=29$ is $1.123R$.
Also notice that the deviation tends to decrease
asymptotically as $N$ increases. This deviation is clearly due to the
exchange energy between electrons of same spin. However this exchange
energy becomes smaller in higher angular momentum shells, which
explains the asymptotic behavior of the self-capacitance.
We also did the calculation for a system of
spinless electrons (i.e no exchange interaction), and found that
the capacitance become then precisely centered around its classical
value $R$.

\noindent
In conclusion we have exactly calculated the energy spectrum and
self-capacitance of a spherical quantum dot in the strong
confinement limit where quantum effects are expected to be
predominant. Remarkably, we have found that the semi-classical theory
remains valid on average in this regime.
We have also found that the main deviations  from the semi-classical
result are due to the exchange interaction between electrons,
but that theses deviations do not exceed 25\%.

\vspace{0.5cm}

We are grateful to J. K. Jain, K. K. Likharev, and T. Kawamura for
insighteful conversations and  critical readings of this manuscript.
This work was supported in part by the office of Naval Research
under grant No. N00014-93-1-0880


\newpage
\begin{figure}
{\caption{Ground state interaction energy $E(N)/N$ as a function
of $N$ for the spin unpolarized case.
Notice that it extrapolates to $0$ when $N=1$.}
\label{fig1}
\end{figure}

\begin{figure}
\caption{(a)The chemical potential and (b)the Self Capacitance
of a spherical quantum dot as
a function of $N$ for the spin unpolarized electron system.}
\label{fig2}
\end{figure}

\begin{figure}
\caption{(a)The chemical potential and (b)the Self Capacitance
of a spherical quantum dot as
a function of $N$ for the spin polarized electron system.}
\label{fig3}
\end{figure}

\end{document}